\documentclass[hidelinks, conference]{IEEEtran}
\usepackage[cmex10]{amsmath}
\usepackage{amssymb} 
\usepackage[pdftex]{graphicx}
\usepackage{cite}

\usepackage{xcolor}
\usepackage[colorlinks=true, linkcolor=darkcyan, citecolor=black, urlcolor=darkcyan, pdfborder={0 0 0}]{hyperref}    
\definecolor{darkcyan}{rgb}{0.0, 0.35, 0.55}

\usepackage{fancyhdr}

\fancypagestyle{firstpage}{%
  \fancyhf{} 
  
  \fancyhead[L]{\scriptsize \copyright~ 2023 IEEE. Personal use of this material is permitted.  Permission from IEEE must be obtained for all other uses, in any current or future media, including reprinting/republishing this material for advertising or promotional purposes, creating new collective works, for resale or redistribution to servers or lists, or reuse of any copyrighted component of this work in other works. 
  DOI: 10.1109/DSP58604.2023.10167765}%
}

\begin{document}
%
\title{Low-Complexity Memoryless Linearizer for Analog-to-Digital Interfaces}

\author{\IEEEauthorblockN{Deijany Rodriguez Linares}
\IEEEauthorblockA{Linköping University\\
	Department of Electrical Engineering\\
	581  83 Linköping, Sweden\\
Email: deijany.rodriguez.linares@liu.se}\\

\and
\IEEEauthorblockN{H\aa kan Johansson}
\IEEEauthorblockA{Linköping University\\
Department of Electrical Engineering\\
581  83 Linköping, Sweden\\
Email: hakan.johansson@liu.se}\\
}


%


\maketitle
\thispagestyle{firstpage}

\begin{abstract}
	This paper introduces a low-complexity memoryless linearizer for suppression
	of distortion in analog-to-digital interfaces. It is inspired by neural
	networks, but has a substantially lower complexity than the neural-network
	schemes that have appeared earlier in the literature in this context.
	The paper demonstrates that the proposed linearizer can outperform
	the conventional parallel memoryless Hammerstein linearizer even when
	the nonlinearities have been generated through a memoryless polynomial
	model. Further, a design procedure is proposed in which the linearizer
	parameters are obtained through matrix inversion. Thereby, the costly and time consuming
	numerical optimization that is traditionally used when training neural
	networks is eliminated. Moreover, the design and evaluation incorporate
	a large set of multi-tone signals covering the first Nyquist band. Simulations show signal-to-noise-and-distortion ratio (SNDR) improvements of some 25 dB for multi-tone signals that correspond to the quadrature parts of OFDM signals with QPSK modulation.
\end{abstract}



%
\IEEEpeerreviewmaketitle

\section{Introduction\label{sec:Intro}}

Conversions between analog and digital signals are essential as the
physical world is analog by nature whereas signal processing is primarily
carried out in the digital domain. These transformations are implemented
using analog-to-digital converters (ADCs) and digital-to-analog converters
(DACs), as well as other components required before/after the ADC/DAC,
like filters, power amplifiers, and mixers. The overall converters
are referred to as analog-to-digital and digital-to-analog interfaces
(ADIs and DAIs), and this paper focuses on ADIs.

With the ever increasing demands for high-performance and low-cost
signal processing and communication systems, there is a need to develop
ADIs with higher performance in terms of data rates (related to bandwidths),
effective resolution [effective number of bits (ENOB) determined by the signal-to-distortion-and-noise ratio (SNDR)\footnote{For a full-scale sinusoidal, $\text{ENOB}=(\text{SNDR}+4.77+P_x)/6.02$ where $P_x$ is the signal power in dB.}],
and low implementation cost (small chip area and low energy consumption).
This project focuses on the SNDR which in analog circuits is degraded
by nonlinearities. The SNDR can consequently be increased by using
linearization techniques which are required to reach a large SNDR.
However, linearization can also be utilized for low/medium SNDRs as
it enables the use of one- or few-bits ADCs, below the targeted SNDR
which is subsequently achieved through digital processing. Thereby
the energy consumption of the ADCs can be reduced substantially as
their energy consumption is very high for large SNDRs, especially
when the data rate is also high \cite{Murmann_2021a}. For example,
the 14-bit converter AD9699 from Analog Devices \cite{AD} has a power
consumption of about 2 W at a data rate of 3 GS/s. It is particularly
important to reduce the ADC power in applications requiring many ADCs
like in massive-MIMO communication systems \cite{Marzetta_2016}.
However, to reach the targeted SNDR, keeping the overall energy consumption
low, it is then vital to develop energy efficient digital linearizers. 

In a practical ADI, the effective resolution is degraded by nonlinearities in the
analog circuitry which introduce distortion, commonly measured in
terms of harmonic distortion (HD) and intermodulation distortion (IMD),
as well as higher-order distortion terms. If the distortion is frequency-independent,
one can use memoryless polynomial modeling and linearization, which
is in focus in this paper. This model is typically sufficient for
narrow to medium bandwidths and resolutions. To reach higher resolutions
over wider frequency bands, one may need to incorporate memory (subfilters)
in the modeling and linearization.

\subsection{Contribution of the Paper and Relation to Previous Work}

In the conventional memoryless linearizer, the distortion is suppressed
by using a special case of a parallel Hammerstein system (Fig. \ref{Flo:Hammerstein-scheme}
in Section \ref{sec:Signal-model-and-proposed-linearizer}) \cite{Chen_95}. This
system incorporates $K$ branches, realizing the terms $c_{k}v^{k}(n)$,
$k=1,2,\ldots,K$, where $v(n)$ is the distorted signal. The implementation
cost of such a scheme emanates from both the $K$ multiplications
between $c_{k}$ and $v^{k}(n)$ and the generation of the $K-1$
nonlinearity terms $v^{k}(n)$. 

In this paper, we introduce a memoryless
linearizer (Fig. \ref{Flo:proposed-scheme} in Section \ref{sec:Signal-model-and-proposed-linearizer})
where the nonlinearity terms $v^{p}(n)$ are replaced by the simpler
nonlinearities $|v(n)|$ or $f_{m}(v)=\max\{0,v(n)\}$ (rectified linear
unit) as they can be implemented in hardware
with low complexity. In addition, bias values are added before the nonlinearities.
The proposed linearizer is inspired by neural networks, but
the neural-network schemes that have appeared earlier in the literature
in this context have a very high implementation complexity (several
hundreds of multiplications in \cite{DENG_202063} and several thousands
of multiplications in \cite{Peng_2021}). The implementation complexity
of the proposed linearizer is substantially lower and it will be demonstrated
that it can outperform the conventional parallel Hammerstein linearizer
even when the nonlinearities have been generated through a memoryless
polynomial model. 

Further, a design procedure is proposed in which
the parameters (corresponding to the multipliers $c_{k}$ above) are
obtained through matrix inversion. Thereby, one can eliminate
the costly and time consuming numerical optimization that is traditionally used when training
neural networks \cite{DENG_202063,Peng_2021}. Moreover, the design
and evaluation incorporate a large set of multi-tone
signals covering the first Nyquist band. In the previous work \cite{DENG_202063,Peng_2021},
the evaluation has only included a few single-tone signals. Our simulations
show SNDR improvements up to some 25 dB for multi-tone 
signals that correspond to the quadrature parts of OFDM signals with QPSK modulation.

Finally, it is noted that neural networks have also been considered for predistortion
of power amplifiers (PAs) required in DAIs, see for example \cite{Tarver_2019,Liu_2022}
and references therein. However, those techniques are not directly
applicable for ADI linearization which is in focus here. A major difference is that, for ADIs, the nonlinearities can be undersampled
and still suppressed provided the signal is Nyquist sampled \cite{Tsimbinos_98}.
In DAIs, the predistorted signal must be oversampled to enable distortion
cancellation in the analog domain.

\subsection{Outline}
Following this introduction, the signal model, Hammerstein linearizer, and proposed linearizer are presented in Section II. Section III presents the proposed design procedure, whereas simulations, results, and comparisons are provided in Section IV. Finally, Section V concludes the paper.

\section{Signal Model, Hammerstein Linearizer, and Proposed Linearizer \label{sec:Signal-model-and-proposed-linearizer}}

\begin{figure}
	\begin{centering}
		\includegraphics{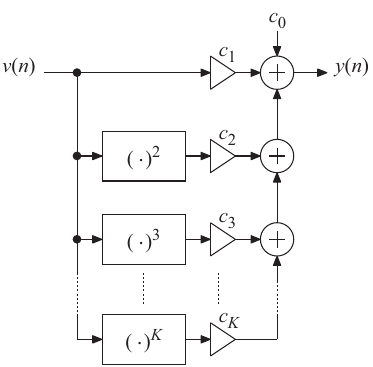}
		\par\end{centering}
	\caption{Hammerstein linearizer.}	
	\label{Flo:Hammerstein-scheme}
\end{figure}

Let the desired discrete-time signal be $x(n)=x_{a}(nT)$ which corresponds
to an analog signal $x_{a}(t)$ sampled uniformly with a sampling frequency of
$1/T$. In addition, the signal contains noise but this is not included
in the mathematical expressions below for the sake of simplicity.
It is further assumed that the distorted signal, say $v(n)$, can be modeled as
a memoryless polynomial according to
\begin{equation}
	v(n)=a_{0}+a_{1}x(n)+\sum_{p=2}^{P}a_{p}x^{p}(n),
	\label{eq:signal-model}
\end{equation}
where $a_{0}$ is a constant, $a_{1}$ is a linear-distortion constant,
and $a_{k},k=2,3,\ldots,P$, are $P-1$ nonlinear-distortion constants.
In the conventional parallel Hammerstein linearizer, depicted in Fig.
\ref{Flo:Hammerstein-scheme}, the linearized signal, say $y(n)$,
is formed as
\begin{equation}
	y(n)=c_{0}+c_{1}v(n)+\sum_{k=2}^{K}c_{k}v^{k}(n).
\end{equation}
In an implementation, this scheme requires $2K-1$ multiplications and $K$
two-input additions per sample ($K$ multiplications for generating $c_{k}v^{k}(n)$,
and $K-1$ multiplications for generating $v^{k}(n)$).

Before proceeding, it is noted that, to be able to model the signal
as in \eqref{eq:signal-model} in a practical system, the parameters $a_{k}$ need to be
estimated. There are several methods available for this purpose \cite{Chen_95}.
However, the focus of this paper is to linearize the distorted signal,
and the model in \eqref{eq:signal-model} is then used for generating
sets of test and evaluation signals. It is stressed though that the
proposed linearizer (to be described below) does not assume that the distorted
signal is in the form of the polynomial in \eqref{eq:signal-model}.

\subsection{Proposed Linearizer \label{sec:Proposed-linearizer}}
\begin{figure}
	\begin{centering}
		\includegraphics{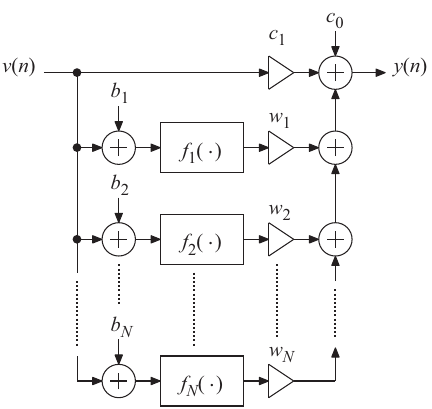}
		\par\end{centering}
	\caption{Proposed linearizer.}	
	\label{Flo:proposed-scheme}
\end{figure}

Given a distorted signal $v(n)$, the output signal $y(n)$ in the proposed linearizer, depicted in Fig. \ref{Flo:proposed-scheme}, is formed as
\begin{equation}
	y(n)=c_{0}+c_{1}v(n)+\sum_{m=1}^{N}w_{m}u_{m}(n)\label{eq:prop_linearizer}
\end{equation}
where
\begin{equation}
	u_{m}(n)=f_{m}(v(n)+b_{m}),
\end{equation}
with $f_{m}$ being nonlinear operations. Here, $f_{m}(v)=|v|$ or $f_{m}(v)=\max\{0,v\}$ due to their low hardware implementation complexity, as noted in \cite{Tarver_2019} where $f_{m}(v)=\max\{0,v\}$ was utilized in a predistorter. Further, the bias values $b_{m}$, $m=1,2,\ldots,N$, are assumed to be uniformly
distributed between $-b_{\max}$ and $b_{\max}$, where the value
of $b_{\max}$ is optimized as outlined in Section \ref{sec:Design}. That is, $b_{m}$
are given by
\begin{equation}
	b_{m}=-b_{\max}+\frac{2(m-1)b_{\max}}{N-1}\label{eq:bm}
\end{equation}

In an implementation, the proposed scheme requires only $N+1$ multiplications per sample with $N$ nonlinear branches, and $2N+1$ two-input additions due to the additional $N$ bias additions. An additional advantage of the proposed linearizer is that it does not require any quantizations before the multiplications. In the Hammerstein linearizer, on the other hand, quantizations are required at the outputs of the nonlinearities (i.e., before the multiplications) to avoid very long and costly internal wordlengths. This introduces quantization errors (quantization noise) which are scaled by the coefficients $c_k$. Therefore, those coefficients have to be kept small to avoid a large noise amplification which would degrade the output SNDR. Alternatively, longer internal wordlengths are needed but it also increases the implementation cost. This problem is circumvented by using the proposed scheme where all quantizations take place after all multiplications. Hence, there is no noise amplification from the quantizations to the output in the proposed scheme.

\section{Design\label{sec:Design}}

The proposed linearizer is designed as described below.
\begin{enumerate}
	\item Form a set of $R$ reference signals $x_{r}(n)$ and the corresponding
	distorted test signals $v_{r}(n)$, $r=1,2,\ldots,R$, using the signal model
	in \eqref{eq:signal-model}.
	\item Form a set of $Q$ uniformly distributed values of $b_{\max}\in[0.5,1]$,
	i.e., $b_{\max}=0.5+0.5q/(Q-1)$, $q=0,1,\ldots Q-1$.
	\item For each value of $b_{\max}$, compute $b_{m}$ according to \eqref{eq:bm}.
	Then, minimize the cost function $E$ as given by
	\begin{equation}
		E=\sum_{r=1}^{R}\sum_{n=1}^{L}\big(y_{r}(n)-x_{r}(n)\big)^{2},\label{eq:E}
	\end{equation}
	where $L$ is the data length. All coefficients $c_{0}$, $c_{1}$,
	and $w_{m}$, $m=1,2,\ldots,N$, in \eqref{eq:prop_linearizer} are
	computed via a matrix inversion, including $\ell_{2}$-regularization
	to avoid ill-conditioned matrices. To be precise,
	let ${\bf w}$ be a $(N+2) \times 1$ column vector containing all
	coefficients $w_{m}$, $m=1,2,\ldots,N$, $c_{1}$, and $c_{0}$,
	and ${\bf A}_{r}$ be an $L\times(N+2)$ matrix where the columns
	$l$, $l=1,2,\ldots,N+2$, contain the $L$ samples $u_{rm}(n)$,
	$m=1,2,\ldots,N$, $L$ input samples $v_{r}(n), $and $L$ ones (for
	the constant $c_0$). Then, minimizing $E$ in \eqref{eq:E} in the least-squares
	sense gives the solution
	\begin{equation}
		{\bf w}={\bf A}^{-1}{\bf b},
	\end{equation}
	where
	\begin{equation}
		{\bf A}=\lambda{\bf I}+\sum_{r=1}^{R}{\bf A}_{r}^{\top}{\bf A}_{r},\quad{\bf b}=\sum_{r=1}^{R}{\bf A}_{r}^{\top} {\bf b}_{r},
	\end{equation}
	with ${\bf A}_{r}^{\top}$ being the conjugate transpose of ${\bf A}_{r}$
	and ${\bf b_{r}}$ being an $L\times1$ column vector containing
	the $L$ samples $x_{r}(n)-v_{r}(n)$.\textbf{ }It is noted here that
	$x_{r}(n)-v_{r}(n)$ is used in ${\bf b_{r}}$ instead of $x_{r}(n)$,
	in order to compute a small value of $c_{1}$. That is, we are replacing
	$c_{1}v(n)$ in \eqref{eq:prop_linearizer} with $v(n)+\Delta c_{1}v(n)$
	and then compute the value of $\Delta c_{1}$. In this way, all parameters
	to be computed in the least squares design are then small (zero in
	the ideal case with no distortion). Further, $\lambda {\bf I}$ is a diagonal
	matrix with small diagonal entries $\lambda$ for the $\ell_{2}$-regularization.
	The compensated channel output $y_{r}(n)$ can be computed as
	\begin{equation}
		y_{r}(n)=v_{r}(n)+{\bf w}^{\top}{\bf A}_{r}^{\top}.
	\end{equation}
	\item Select the best of the $Q$ solutions above.
	\item Evaluate the linearizer over a large set of signals, say $M$ signals, where $M \gg R$.
\end{enumerate}

\section{Simulations, Results, and Comparisons}
We assume a distorted signal $v(n)$ as in \eqref{eq:signal-model} with the parameters\footnote{We have also considered the case with $a_0 \approx 0$ and $a_1 \approx 1$ and obtained roughly the same results.} $a_0=0$, $a_1=1$, $a_p=(-1)^p\times0.15/p$, $p=2,3,\ldots,P$, and $P=10$. The signal $x(n)$ is selected as
\begin{equation}
	x(n)=\frac{0.9}{31}\times \sum_{k=1}^{31} A_k\sin (\omega_k n+\alpha_k),
\end{equation}
where 
\begin{equation}
w_k=\frac{2\pi k}{64}+\Delta\omega.
\end{equation}
This signal corresponds to the quadrature part (imaginary part) of an OFDM signal with 31 active carriers out of 64 carriers and frequency offset represented by $\Delta\omega$. Further, we assume QPSK modulation which means that $A_k=1$ for all $k$ whereas $\alpha_k$ are randomly generated from the four possible angles $\pi/4$, $-\pi/4$, $3\pi/4$, and $-3\pi/4$. The scaling constant 0.9 is included to ensure that the magnitude of the distorted signal is less than one.

In the design, we use only one such signal (thus $R=1$ in Section \ref{sec:Design}), with all $\alpha_k$ being zero but with a random frequency offset. In the evaluation, we use $M=500$ signals with randomly generated frequency offsets assuming uniform distribution between $-\pi/64$ and $\pi/64$. All signals are of length $L=2^{13}$ and quantized to $12$ bits. Further, we have considered both $f_{m}(v)=|v|$ and $f_{m}(v)=\max\{0,v\}$ for the nonlinear operations. The results below are for $f_{m}(v)=|v|$ but the other option gave roughly the same results. Fo the $\ell_{2}$-regularization, we have used $\lambda=0.02$.

\begin{figure}
	\begin{centering}
		\includegraphics[scale=0.4]{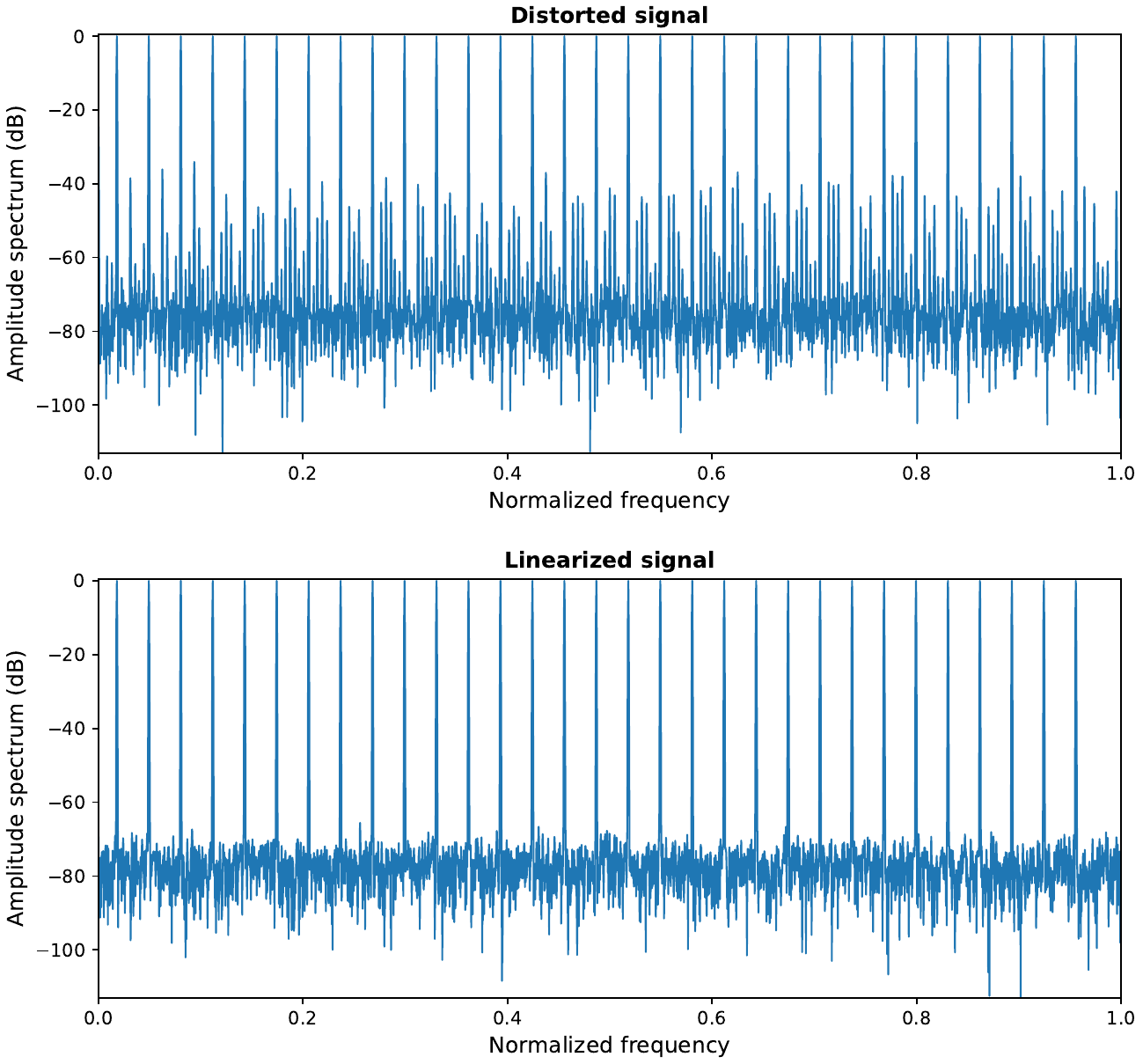}
		\par\end{centering}
	\caption{Spectrum before and after linearization for one of the signals using the proposed linearizer with $N=16$ nonlinear branches.}	
	\label{Flo:Spectrum}
\end{figure}

\begin{figure}[b]
	\begin{centering}
		\includegraphics[scale=0.4]{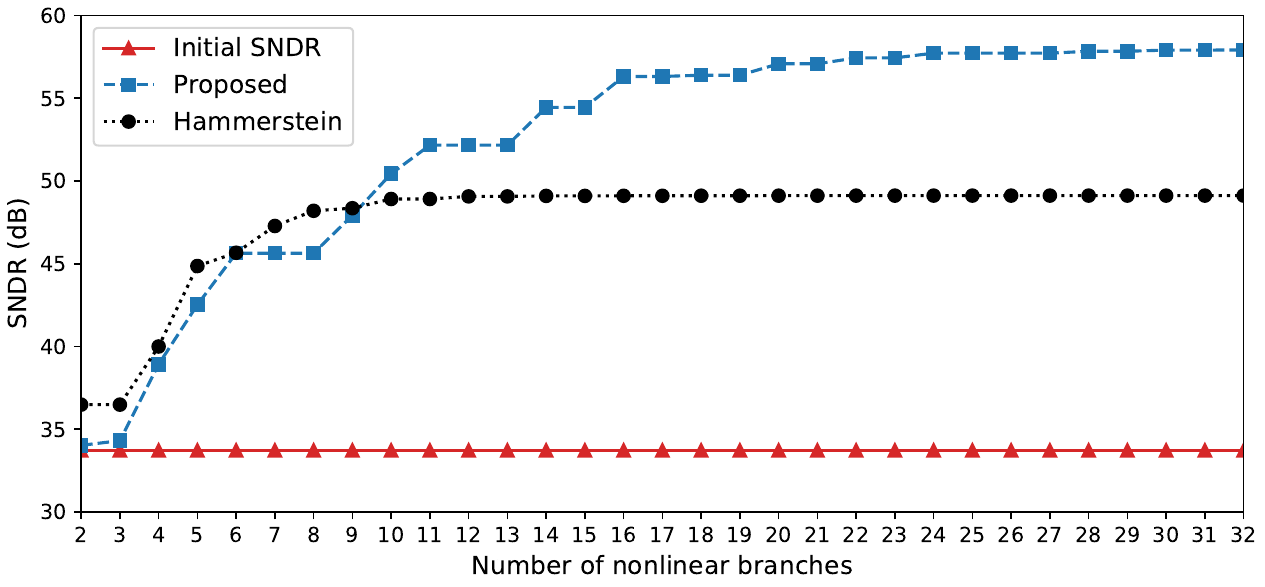}
		\par\end{centering}
	\caption{SNDR versus number of nonlinearity branches.}	
	\label{Flo:SNDR_versus_numb_branches}
\end{figure}

\begin{figure}
	\begin{centering}
		\includegraphics[scale=0.4]{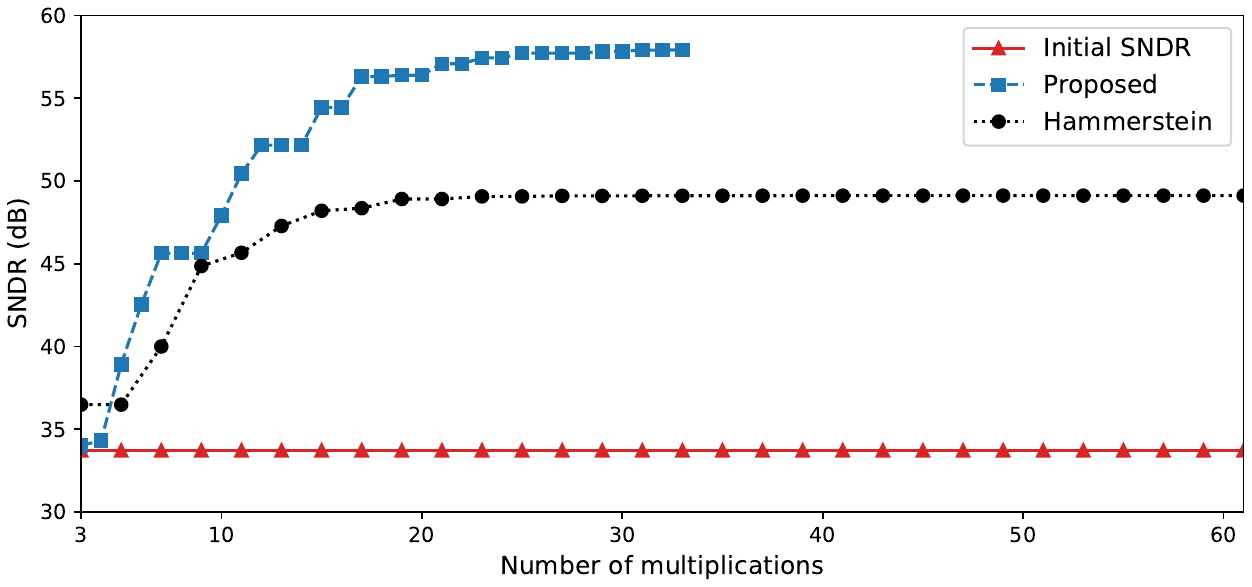}
		\par\end{centering}
	\caption{SNDR versus number of multiplications.}	
	\label{Flo:SNDR_versus_multiplications}
\end{figure}

Figure \ref{Flo:Spectrum} plots the spectrum before and after linearization for one of the signals. Figure \ref{Flo:SNDR_versus_numb_branches} plots the SNDR using both the proposed linearizer and the Hammerstein linearizer, which has been designed in the same way as the proposed linearizer but without the bias values. When comparing the implementation complexities for the same SNDR, one should however not use the number of branches but rather the number of multiplications and additions required, of which the multiplications are the most expensive ones to implement. Recall from Section \ref{sec:Signal-model-and-proposed-linearizer} that the Hammerstein linearizer requires $2K-1$ multiplications for $K-1$ nonlinear branches whereas the proposed linearizer requires only $N+1$ multiplications for $N$ nonlinear branches. To take this into account, Fig. \ref{Flo:SNDR_versus_multiplications} plots the SNDR versus the multiplication complexity.

As seen in Fig. \ref{Flo:SNDR_versus_multiplications}, the proposed linearizer outperforms the Hammerstein linearizer when the number of of multiplications exceeds 4. Furthermore, it is noted that for the signal considered in this example, quantized to 12 bits, the SNR is approximately 58 dB when there is no distortion. As seen in Fig. \ref{Flo:SNDR_versus_numb_branches} the SNDR approaches this value for the proposed linearizer when the number of nonlinear branches increases. With about 20 nonlinearity branches, the SNDR is close to 58 dB. For the Hammerstein linearizer, the SNDR saturates at a lower level. This is due to the  $\ell_{2}$-regularization which is needed to keep the nonlinearity multiplier values $c_k$ small to avoid large quantization noise amplification, as discussed in Section \ref{sec:Proposed-linearizer}. With $\lambda=0.02$, used in this example, all multiplier values are below some 1/2 in magnitude which means that only a few additional bits are needed internally in the Hammerstein linearizer. As also discussed in Section \ref{sec:Proposed-linearizer}, such quantization amplification does not exist in the proposed linearizer, but also in this linearizer all nonlinearity multiplier values are below some 1/2 in magnitude.

\section{Conclusion}
This paper introduced a low-complexity memoryless linearizer for suppression
of distortion in analog-to-digital interfaces. It is inspired by neural
networks, but has a substantially lower complexity than traditional neural-network schemes (as seen in Fig. \ref{Flo:SNDR_versus_multiplications}). Simulations demonstrated SNDR improvements of some 25 dB for multi-tone signals that correspond to the quadrature parts of OFDM signals with QPSK modulation. Thereby, the SNDR approached the SNR of such multi-tone signals quantized to 12 bits. It was also demonstrated that the proposed linearizer can outperform the conventional parallel memoryless Hammerstein linearizer, both in terms of low implementation complexity for the same SNDR, and in terms of feasible SNDR improvements. 

Furthermore, a design procedure has been proposed in which all linearizer parameters are obtained through matrix inversion, thereby avoiding the costly and time-consuming numerical optimization that is traditionally used when training neural networks. Ongoing work studies the extension to low-complexity memory linearizers addressing frequency-dependent models.

\begin{small} 
	\bibliographystyle{IEEEtran}
	\bibliography{bibliography}
	
\end{small}
\end{document}